\documentstyle[psfig]{mn}

\input{epsf}

\title[Velocity dispersion estimates of APM galaxy clusters]
{Velocity dispersions estimates of APM galaxy clusters\thanks{Based 
on observations collected at CASLEO Observatory.  Complejo Astron\'omico
El Leoncito is operated under agreement between the CONICET, Argentina and
the National Universities of La Plata, C\'ordoba and San Juan.}}

\author[M. Victoria Alonso et al.]{M. Victoria Alonso,
 Carlos Valotto \thanks {On a fellowship from CONICET, Argentina}, Diego G. Lambas, and
Hern\'an Muriel\\
Grupo de Investigaciones en Astronom\'\i a Te\'orica y Experimental 
(IATE), Observatorio Astron\'omico de C\'ordoba, \\
Laprida 854 (5000), C\'ordoba, Argentina and Consejo Nacional de Investigaciones Cient\'\i ficas y T\'ecnicas (CONICET), Argentina}

\date{\today}

\begin{document}
 
\maketitle
 
\begin{abstract}

We present 83 new galaxy radial velocities in the field of 18 
APM clusters with 
redshifts between 0.06 and 0.13.
 The clusters have Abell identifications 
and the galaxies were selected within 0.75 h$^{-1}$Mpc in projection from their  
centers.
We derive new cluster velocity dispersions for 13
clusters using our data and
published radial velocities.

We analyze correlations between cluster velocity 
dispersions and  
cluster richness counts as defined in Abell and APM catalogs. 
The correlations
show a statistically significant trend although with a large scatter
suggesting that richness is a  poor estimator of
cluster mass irrespectively of cluster selection criteria and richness
definition. 
We find systematically lower velocity dispersions in the sample
of Abell clusters that do not fulfill APM cluster selection criteria suggesting
artificially higher Abell richness counts due to  
contamination by projection effects in this subsample.

\end{abstract}


\section{INTRODUCTION}

Studies of the dynamics of clusters of galaxies play an important role
in the analysis of large scale structure formation.  
Cluster velocity dispersion 
measurements $\sigma$ provide cluster mass estimates and 
a direct normalization of the primordial mass power spectrum (see for
instance Eke et al. 1996). 
Samples of Abell clusters have been extensively used in these analyses. 
However, studies of selection effects in the Abell catalog (Sutherland 1988, 
Dalton 1992) have shown the presence of serious projection effects
and plate calibration systematics. 
On the other hand, numerical simulations \cite{vanhaarlem} provide 
 evidence that cluster surveys in two dimensions are subject to strong projection
biases if the cluster search radius is as large as Abell's radius R$_A$=1.5 h$^{-1}$ Mpc.
Thus, clusters selected with this criteria are subject to a frequent superposition of groups that 
may produce artificial large velocity dispersions. Nevertheless, 
these authors find that clusters obtained from two dimensional analysis but with a 
significantly
smaller search radius, R=0.5 h$^{-1}$ Mpc, have similar distributions of velocity
dispersions than those clusters selected in three dimensions. 
Several studies  based on different observational 
samples Frenk et al. (1990), Girardi et al. (1993), Zabludoff et al. (1993),
Collins et al. (1995), Mazure et al. (1996), Fadda et al. (1996) have
provided insights on the kinematics of galaxies in clusters. 
These works have been based on Abell clusters where spurious high velocity
dispersions may be expected due to projection effects.

The Edinburgh--Durham cluster catalog
(Lumsden et al. 1992) although free from subjective visual systematics, 
would also be biased 
toward artificial large velocity dispersions due to 
superpositions given that the same search radius 
than in Abell's catalog is used in the cluster identifications. 
Other automated survey, the APM cluster
catalog (Dalton et al. 1994, 1997) has an intermediate search radius 0.75 h$^{-1}$ Mpc although galaxies 
in the outer ring
0.50--0.75 h$^{-1}$ Mpc have a smaller weight in the calculation of cluster richness.
Thus, it might be expected 
that the distribution of APM cluster velocity dispersions would be
more representative of the true distribution. There are 31 APM identifications 
(Mazure et al. 1996) in the ESO Nearby Abell Cluster Survey (ENACS).  In a quantitative 
analysis of these data the authors conclude that 
the large spread between velocity dispersion and richness, both APM and Abell,
is probably or at least partially intrinsic to the clusters. 

In order to improve the sample of APM clusters with velocity dispersion  
estimates we have undergone an observational
program to obtain radial velocities of galaxies in the field of APM clusters.
We present in this paper 
new measurements of radial velocities of galaxies in the fields
of 18 APM clusters.  Our data combined with radial velocities 
from the literature allow us to determine cluster velocity dispersions
for 17 APM clusters (13 of these without previous estimates).
In section 2 we present the galaxy data set and a statistical analysis 
of the new velocity dispersion estimates and those from the literature
providing correlations between $\sigma$ and richness counts $C$. 
A brief discussion of the results is given in section 3.

\section{DATA AND ANALYSIS}

We aim to estimate velocity dispersions of APM clusters for a wide range of richness.
Therefore, we have selected a sample of APM 
clusters (Dalton et al. 1994, 1997, hereafter APM IV and APM V respectively) 
with redshifts between 0.06 and 0.13 and uniformly distributed in richness.
To avoid confusion, cluster names are as in 
NASA/IPAC Extragalactic Database
(NED). 
Our sample comprises 23 APM clusters for which we have selected galaxies with
 $b_j < $ 19.5 
from the Edinburgh--Durham Southern Galaxy Catalogue
 (Heydon-Dumbleton
et al. 1989, hereafter COSMOS)
 within 0.75 Mpc $h^{-1}$ in projection from the cluster centers 
(Dalton et al. 1994).

We have chosen the APM clusters in our sample to have Abell identifications
(Abell 1958, Abell et al. 1989) in order to perform a comparative analysis
between the cluster dynamics and richness. 
In Table 1 we show basic information of our original sample.
We list in column 1 and 2, 
the cluster identifications (APM and Abell respectively); columns 3 and 4, coordinates 
of cluster centers;
columns 5 and 6, mean cluster redshift and the number of objects used in these 
calculations from Ebeling \& Maddox (1995); 
and columns  7 and 8, the richness parameter from Abell and APM V.
As it can be seen from the Table, most of the mean cluster redshifts are based on
 measurements of two members
and it is very important to improve these cluster redshifts.

The spectroscopic observations were carried out during 1996 and 1997 
using a REOSC spectrograph in the 2.15 m telescope at CASLEO Observatory, Argentina.
We have used a 600 line mm$^{-1}$ grating with a resolution of 3.3$\AA$.  We observed
the galaxies twice with typical exposure times of about 20 minutes to avoid collecting
many cosmic rays.  The spectral range 
was 4000$\AA$ to 7500$\AA$ and the spectra were calibrated using comparison lines from
a He--Ne--Ar lamps with an accuracy of 15 km s$^{-1}$.  We also observed galaxies with known
radial velocities to be used as templates. 

The data reductions were performed using the standard procedure to remove bias
images, correct by flat-field and make illumination corrections using IRAF 
routines.  Radial velocities were obtained following the
cross--correlation method of Tonry \& Davies \shortcite{tonry}.  

Table 2 shows radial velocities obtained with galaxies in common with other authors,
listing the identification, coordinates, $b_j$ magnitude, our heliocentric radial 
velocities and
the measurements from other authors, respectively.  Our measurements are in
 good agreement with those from literature, with a mean difference  
of 52$\pm$60 km s$^{-1}$, lesser than quoted errors. 

We provide in Table 3 our new radial velocity measurements of galaxies in the
fields of 
our selected APM clusters.  For each cluster, column 1 lists galaxy identification,
using names taken from the Guide Star Catalog \cite{lasker}, APM (Maddox et al. 1990a,
Maddox et al. 1990b),
APMBGC \cite{loveday} or the catalogue of principal galaxies (Paturel 1989, PGC) 
whenever available; 
columns 2 and 3, the equatorial coordinates; column 4, $b_j$ magnitude when available; 
column 5, the observed
heliocentric radial velocity, V$_r$ and the associated standard deviation.  Those galaxies
marked with an asterisk in column 2 are not in COSMOS Survey. Quoted coordinates 
are from our own identification.

We have also searched for available redshifts in the area of APM clusters using
the NASA/IPAC Extragalactic Database in order to improve our $\sigma$ estimates.  
We have identified 
52 galaxies from The Las Campanas Redshift Survey \cite{schectman} within
0.75h$^{-1}$ Mpc in projection in the fields of the 
clusters APMCC 160, APMCC 173, APMCC 352, APMCC 746, APMCC 042 and APM 221539.0-390817.
 
Based on the  ROSTAT routine (see Beers et al. 1990) 
we have used robust
mean and scale estimators. We have applied relativistic corrections and
we have taken into account velocity errors. Considering the typical
number of redshift confirmed cluster members (usually $<$ 20)
we have considered the {\it trimean}
estimator for the mean velocity and the {\it gapper} for the velocity
dispersion. Errors are based on the statistical {\it jacknife}.

When possible, we have analyzed the velocity and the projected
distributions in order to detect subclustering. In the cluster APMCC 746 
we find a substructure separated from the main cluster in both radial velocity and
projected coordinates. This structure was previously identify as the group of galaxies 
AM 2159-224 (Arp and Madore, 1987). We obtained a mean radial velocity for this group $V=21124$ km/s, with  
a difference of 476 km/s with respect to the main cluster. 
Thus,  mean radial velocity and $\sigma$ of cluster APMCC 746 were
computed after removing this structure.

We have not computed velocity dispersions for several clusters in our sample.
APMCC 107 has several substructures and more redshifts are needed to derive an
accurate velocity dispersion. The clusters: APM 032010.5-454456, APMCC 604 and 
APMCC 864 have few redshift
measurement to estimates the velocity dispersion.

Finally, we have computed new velocity 
dispersion estimates for 13 clusters. In 11 clusters, the estimates rely on
our new radial velocity measurements, and in 2 of them on the Las Campanas Redshift
Survey (Schectman et al. 1996).
In Table 4 we list cluster identification; our new estimates of mean cluster redshift and
velocity dispersions; and the number of objects used in these
measurements. The mean redshifts, given in this Table, provide a more confident source of APM
cluster redshifts given a larger number of members.  
There is a general good agreement between the cluster mean redshifts and the estimates 
quoted in Ebeling \& Maddox (1995).
Redshift uncertainties are smaller than 0.001 and are not quoted in the Table.

We also present velocity dispersions for four APM clusters in common with  Fadda et al.
\shortcite{fadda}, the largest survey
for determination of cluster velocity dispersions.  
Table 5 shows our results for 
these clusters listing cluster 
identifications, our new estimate for the
cluster redshift, our velocity dispersion estimates and those from Fadda et 
al. (1996). 
The mean difference between our results and  Fadda et al. \shortcite{fadda}
 is -63$\pm$50 showing a good
agreement in spite of the small number
of member galaxies in our analysis. 

\begin{figure}
{\psfig{file=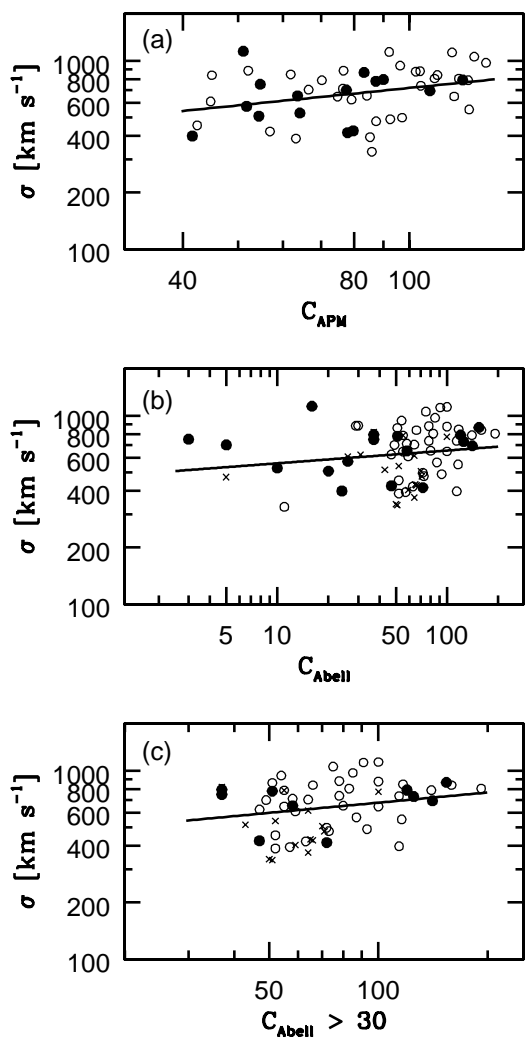,height=15cm}}                         
\caption{Correlations between velocity dispersions of clusters and richness.
Panel 1a corresponds to APM richness counts,
solid circles to our new velocity dispersions and open circles to those
obtained from the literature. Panel 1b and 1c correspond to
Abell richness count, where only richnnes class $R>0$ are considered in 1c. 
Crosses identify Abell-non
APM clusters in the same survey area and redshift range than the APM survey.
}
\label{fig:fig1}
\end{figure}
 
We have added to our new velocity dispersion determinations, estimates taken  
from the literature (Mazure et al. 1996, Fadda et al. 1996, Struble and Rood 1991,
Zabludoff et al. 1993, and Muriel
et al. 1999). 
We consider APM richness counts
as defined in APM V and we analyze  
a sample of 48 APM clusters with reliable velocity dispersions.
In order to compare the correlations of APM and Abell
richness counts with velocity dispersions we have added to the previous
compilation of APM/Abell clusters, those  
Abell clusters  within the same area
and redshifts of the APM cluster survey that were not identified by the
APM selection criteria.

Figure 1 shows the correlation between $\sigma$ and $C$. 
Figure 1a corresponds to APM richness counts $C_{APM}$,
with solid circles representing our new velocity dispersions and open circles those
obtained from the literature. Figure 1b and 1c correspond to
Abell richness count $C_{Abell}$, where only richnnes class $R>0$ are considered in 1c. 
In Figures 1b and 1c
are also shown as crosses Abell clusters    
in the same survey area and redshift range than the APM survey that
did not fulfill the APM V cluster selection criteria.
In these Figures there is a clear tendency of these clusters
to have systematically lower values of velocity dispersion compared to
those clusters in the APM survey, suggesting that "non-APM" clusters have artificially
larger Abell richness counts.   

In Figure 1a, b and c are also shown the corresponding 
least-squares power-law fits. 
The values of the best fitting parameters  of the form  
$log(\sigma)=a \times log(C) + b $ for the correlations shown in this
Figure are: APM a=0.307$\pm$0.134, b=2.242$\pm$0.256;
Abell a=0.068$\pm$0.048 ,b=2.680$\pm$0.084 and  
Abell ($C_{Abell} > 30$) a=0.180$\pm$0.097 ,b=2.471 $\pm$0.180.  
The corresponding rms scatter around these
fits are 0.016, 0.018 and 0.017 respectively. 
It can be appreciated a poor
correlation between $\sigma$ and richness counts 
irrespective of the different procedures indicating that
this scatter is partially intrinsic to the clusters. 
This should be taken into account when deriving  mass estimates and
abundances from cluster richness (Bahcall and Cen 1993), which could introduce 
spurious biasing effects against low $\sigma$ objects as discussed by Mazure et al. (1996).

\section{DISCUSSION}

The measurements of galaxy redshifts in the field of
APM clusters contribute to our understanding of
the dynamics of these systems and provide a deeper insight
in the problem of cluster identification from
projected data.
The estimated velocity dispersions
obtained from our analysis with typical number of galaxies
$\simeq 10-20$ are in good agreement with estimates derived
from the literature as seen in Table 5.

The new observations presented in this paper together 
with previous published data 
allows for a comparative analysis of the
correlation between cluster velocity dispersion and richness defined by different 
selection algorithms.
The correlation between Abell cluster richness and velocity dispersions
is  very poor.  
Automated cluster catalogs such as APM are free from
subjective effects which may blur significantly the
correlation between richness and $\sigma$.
However, APM richness counts do not
provide a significantly better correlation with $\sigma$,
suggesting the presence of an intrinsic spread related to
galaxy formation or evolution in clusters.
Therefore, the observed spread in the richness - $\sigma$ correlations 
raise serious concerns on the use of richness in cluster mass determinations 
irrespectively of cluster selection criteria and richness definition.
 
We observe a systematic trend to lower values of $\sigma$ in the sample
of Abell clusters that did not satisfy APM cluster selection criteria. 
This effect may be related to the fact that these objects  
are subject to larger contamination by projection effects as has been previously 
determined (Sutherland 1988, Dalton 1992, Van Haarlem et al. 1997).

\section{Acknowledgments}
The authors are specially grateful to C. Caretta who obtained some spectra for
this project in 1996 and C. N. A. Willmer for his help in reducing the data.
The authors acknowledge use of the CCD and data acquisition system supported under
U.S. National Science Foundation grant AST-90-15827, to R. M. Rich.

This research has made use of the NASA/IPAC Extragalactic Database
(NED) which is operated by the Jet Propulsion Laboratory, California
Institute of Technology, under contract with the National Aeronautics and
Space Administration. 

This research was supported by grants from  CONICET,
CONICOR, Agencia Nacional de Promoci\'on Cient\'{\i}fica y Tecnol\'ogica, Secretar\'{\i}a 
de Ciencia y T\'ecnica
de la Universidad Nacional de C\'ordoba and Fundaci\'on Antorchas, Argentina.

\begin{table*}
\begin{minipage}{80mm}
\caption{APM cluster sample}
\tabskip =1em plus2em minus.5em
\label {table1}
\begin{tabular}{@{}llrlrrrr}
APM Id. & Abell Id. & $\alpha$(2000) & $\delta$(2000) & $<$ z $>$ & N & $C_{Abell}$ &  $C_{APMCC}$  \\
 \noalign {\vskip 10pt}
 \noalign {\hrule}
 \noalign {\vskip 10pt}
APMCC 015 & A2734 & 00 11 30.07 & -28 51 29.74 & 0.062 & 2  & 58  &  63.6  \\   
APMCC 042 & A2755 & 00 17 44.70 & -35 09 30.30 & 0.095 & 3  & 120 & 124.0  \\   
APMCC 050 & A0022 & 00 20 35.10 & -25 41 57.30 & 0.131 &    & 141 & 108.6  \\   
APMCC 073 & A0042 & 00 28 37.72 & -23 36 43.89 & 0.109 &    & 154 &  83.2  \\   
APMCC 107 & A2819 & 00 46 04.10 & -63 35 13.00 & 0.087 & 2  & 90  & 128.9  \\   
APMCC 123 & S0106 & 00 56 24.95 & -37 53 43.71 & 0.118 & 2  &     &  53.7  \\   
APMCC 132 & S0112 & 00 57 56.70 & -66 48 05.60 & 0.067 & 2  & 16  &  51.1  \\   
APMCC 160 & S0144 & 01 17 35.15 & -37 59 55.99 & 0.077 &    & 26  &  51.8 \\   
APMCC 173 & A2911 & 01 26 17.83 & -37 55 25.16 & 0.079 & 2  & 72  &  77.9  \\   
APMCC 352 & A3098 & 03 13 38.60 & -38 18 20.90 & 0.083 & 2  & 38  &  48.2  \\   
APMCC 359 & S0333 & 03 15 32.43 & -29 15 27.50 & 0.067 & 2  & 24  &  41.6  \\   
APM 031451.8-510556 & A3110 & 03 16 23.30 & -50 54 57.20 & 0.075 && 37  &   \\   
APMCC 369 & S0336 & 03 17 39.10 & -44 31 27.10 & 0.076 & 1  & 5   &  77.5 \\   
APM 032010.5-454456 & S0345 & 03 21 51.60 & -45 34 16.20 & 0.069 & 3  &     &   \\   
APMCC 400 & S0356 & 03 29 30.00 & -46 00 32.30 & 0.072 & 2  & 10  &  64.2  \\   
APMCC 604 & A3703 & 20 39 44.50 & -61 13 59.60 & 0.071 &    & 52  &  42.4 \\   
APMCC 746 & S0987 & 22 02 07.60 & -22 35 52.10 & 0.070 & 15 & 20  &  54.4 \\   
APM 221539.0-390817 & A3856 & 22 18 36.22 & -38 53 14.03 & 0.126 & 2  & 125 &   \\ 
APMCC 815 & A3910 & 22 45 55.30 & -45 54 45.70 & 0.091 & 3  & 47  &  79.7 \\   
APMCC 824 & A3922 & 22 49 45.90 & -51 47 58.40 & 0.098 & 2  & 51  &  87.5 \\   
APMCC 864 & S1096 & 23 11 50.23 & -29 03 49.04 & 0.117 & 1  &     &  63.0 \\ 
APMCC 898 & A2599 & 23 26 47.29 & -23 50 59.10 & 0.098 & 2  & 51  &  58.5 \\   
APMCC 915 & S1140 & 23 39 39.20 & -45 59 08.20 & 0.067 & 1  & 3   &  54.7 \\   
\end{tabular}
\end{minipage}
\end{table*}

\begin{table*}
\begin{minipage}{80mm}
\caption{Radial velocities in common with other authors}
\tabskip =1em plus2em minus.5em
\label {table2}
\begin{tabular}{llllrl}
Name  & $\alpha$(2000) & $\delta$(2000) & $b_j$ & V$_r$ & V$_r$(lit) \\
 \noalign {\vskip 10pt}
 \noalign {\hrule}
 \noalign {\vskip 10pt}
APMBGC 409-109-058 & 00 11 55.2 & -28 43 50.3 & 15.85 & 19741$\pm$147 & 19871$^{(7)}$ \\   
B011514.6-381709 & 01 17 31.1 & -38 01 20.0 & 17.38 & 22405$\pm$296 & 22457$\pm$120$^{(2)}$  \\   
             & 01 26 09.9 & -37 56 42.6 & 17.38 & 24334$\pm$209 & 24307$\pm$39$^{(3)}$ \\   
PGC~0012161  & 03 16 31.1 & -50 54 41.0 & 13.88 & 22065$\pm$149  & 22050$^{(4)}$ \\   
APMBGC 248-116+062 & 03 29 52.3 & -46 02 19.0 & 12.43 & 21553$\pm$145 & 21294$\pm$24$^{(5)}$ \\   
             & 22 47 02.4 & -45 52 13.6 & 16.66 & 15125$\pm$171 & 15163$\pm$75$^{(6)}$ \\   
APM 232411.28-240749 & 23 26 49.3 & -23 51 18.0 & 17.23 & 26874$\pm$254 & 26591$^{(1)}$ \\   
\end{tabular}
\medskip
\medskip

{Sources from radial velocities, V$_r$(lit)\\
1. Dalton et al. \shortcite{dalton} \\
2. Schectman et al. (1996) \\
3. Collins et al. (1995) \\
4. Paturel et al. (1995, LEDA)\\
5. Loveday et al. (1996) \\
6. Di Nella et al. (1996) \\
7. Katgert et al. (1998)} \\
\end{minipage}
\end{table*}

\begin{table*}
\begin{minipage}{80mm}
\caption{New radial velocities for galaxy in APM clusters}
\tabskip =1em plus2em minus.5em
\label {table3}
\begin{tabular}{llllr}
Name  & $\alpha$(2000) & $\delta$(2000) & $b_j$ & V$_r$ \\
 \noalign {\vskip 10pt}
 \noalign {\hrule}
 \noalign {\vskip 10pt}
             &             &              &       &         \\
APMCC 015        &             &              &            \\
             &             &              &       &         \\
             & 00 10 24.3 & -28 49 35.1 & 16.75 & 17886$\pm$199         \\   
             & 00 10 32.4 & -28 51 54.2 & 16.85 & 17819$\pm$286         \\   
             & 00 12 04.9 & -28 47 05.3 & 17.15 & 19366$\pm$215         \\   
             & 00 11 35.0$^*$ & -29 01 24.0 &  --   & 17436$\pm$225         \\   
             & 00 11 18.5$^*$ & -28 50 22.0 &  --   & 18188$\pm$428         \\   
             &             &              &       &         \\
APMCC 050    &             &              &       &         \\
             &             &              &       &         \\
             & 00 20 38.9 & -25 35 30.0 & 15.80 & 19351$\pm$169         \\   
             & 00 20 35.0 & -25 39 26.8 & 17.70 & 34254$\pm$181         \\   
             &             &              &       &         \\
APMCC 073    &             &              &       &         \\ 
             &             &              &       &         \\
             & 00 28 51.6 & -23 36 24.9 & 16.40 & 17659$\pm$204         \\   
             & 00 27 53.8 & -23 41 47.6 & 17.50 & 33573$\pm$131         \\   
             & 00 28 59.1 & -23 31 56.3 & 17.50 & 19561$\pm$281         \\   
             & 00 28 11.9 & -23 42 07.8 & 17.60 & 27035$\pm$224         \\   
             &             &              &       &         \\
APMCC 107    &             &              &       &          \\
             &             &              &       &         \\
8844.0365    & 00 45 12.9 & -63 33 13.0 & 15.37 & 22423$\pm$289         \\   
8844.0574    & 00 45 22.1 & -63 37 27.0 & 15.16 & 23492$\pm$395         \\   
8844.0773    & 00 44 56.9 & -63 28 36.0 & 15.05 & 23100$\pm$350         \\   
8845.0436    & 00 46 20.3 & -63 28 06.0 & 14.20 & 25912$\pm$365         \\   
             &             &              &       &         \\
APMCC 123    &             &              &       &         \\ 
             &             &              &       &         \\
             & 00 55 30.7 & -37 49 52.6 & 16.82 & 30759$\pm$558         \\
             &             &              &       &        \\
APMCC 132    &             &              &       &        \\ 
             &             &              &       &         \\
8848.0146    & 00 57 11.9 & -66 43 49.0 & 14.78 & 19949$\pm$280         \\   
8848.0287    & 00 58 20.7 & -66 48 00.0 & 14.77 & 19863$\pm$269         \\   
8848.0376    & 00 57 46.1 & -66 47 50.0 & 15.12 & 18771$\pm$298         \\   
8848.1300    & 00 58 11.9 & -66 48 17.0 & 14.86 & 20850$\pm$309         \\   
             & 00 58 20.0$^*$ & -66 47 58.0 &  --   & 19200$\pm$223         \\ 
             &             &              &       &         \\
APMCC 160    &             &              &       &                  \\
             &             &              &       &         \\
             & 01 17 12.9 & -38 04 17.2 & 17.78 & 23421$\pm$276         \\   
             & 01 17 26.0$^*$ & -38 01 40.0 &  --   & 32923$\pm$298         \\   
             &             &              &       &         \\
APMCC 359    &             &              &       &         \\
             &             &              &       &         \\
             & 03 16 08.9 & -29 18 20.3 & 15.96 & 19354$\pm$127         \\   
             & 03 16 13.9 & -29 13 50.2 & 16.96 & 20109$\pm$181         \\   
             &             &              &       &         \\
APM 031451.8-510556 &             &              &       &         \\
             &             &              &       &         \\
8065.1131    & 03 16 32.1 & -50 54 08.0 & 15.32 & 18293$\pm$227         \\   
8065.1191    & 03 16 37.0 & -50 53 06.0 & 15.23 & 23536$\pm$206         \\   
8065.1317    & 03 15 56.2 & -50 50 11.0 & 15.44 & 16788$\pm$191         \\   
8065.1387    & 03 16 32.8 & -50 58 59.0 & 15.51 & 26934$\pm$219         \\   
             & 03 16 24.3 & -50 52 39.8 & 17.40 & 22196$\pm$141         \\   
             & 03 16 17.5 & -50 56 56.3 & 17.70 & 22470$\pm$171         \\   
             & 03 16 34.6 & -50 54 06.7 & 17.90 & 23108$\pm$215         \\   
             & 03 16 04.5$^*$ & -50 55 46.0 &  --   & 23153$\pm$165         \\ 
             & 03 16 29.0$^*$ & -50 54 08.0 &  --   & 23794$\pm$178         \\ 
             & 03 16 27.0$^*$ & -50 54 41.0 &  --   & 22310$\pm$243         \\ 
\end{tabular}
\end{minipage}
\end{table*}

\begin{table*}
\begin{minipage}{80mm}
\contcaption{}
\tabskip =1em plus2em minus.5em
\begin{tabular}{llllr}
Name  & $\alpha$(2000) & $\delta$(2000) & $b_j$ & V$_r$ \\
\noalign {\vskip 10pt}
\noalign {\hrule}
\noalign {\vskip 10pt}
             &             &              &       &         \\
APMCC 369    &             &              &       &         \\
             &             &              &       &         \\
             & 03 17 17.9 & -44 21 22.2 & 17.17 & 24253$\pm$167         \\   
             & 03 17 01.7 & -44 21 05.1 & 17.57 & 27165$\pm$303         \\   
             &             &              &       &         \\
APM 032010.5-454456 &             &              &       &         \\
             &             &              &       &         \\
             & 03 21 59.1 & -45 33 22.3 & 16.77 & 18418$\pm$157         \\   
             &             &              &       &         \\
APMCC 400    &             &              &       &         \\
             &             &              &       &         \\
8059.0142    & 03 28 58.7 & -45 56 00.0 & 15.03 & 21994$\pm$153         \\   
8059.0478    & 03 29 08.1 & -45 58 24.0 & 14.71 & 20771$\pm$243         \\   
8060.0804    & 03 30 00.4 & -46 05 47.0 & 13.04 & 20219$\pm$176         \\   
             & 03 29 16.7 & -46 04 50.1 & 17.67 & 21503$\pm$273         \\   
             &             &              &       &         \\
APMCC 604    &             &              &       &         \\
             &             &              &       &         \\
9100.0418    & 20 38 42.5 & -61 18 10.0 & 14.25 & 22300$\pm$305         \\   
9100.0429    & 20 39 55.6 & -61 17 38.0 & 14.39 & 21439$\pm$202         \\   
9100.0519    & 20 39 46.9 & -61 11 23.0 & 13.87 & 27392$\pm$199         \\   
9100.0541    & 20 39 03.7 & -61 10 54.0 & 13.40 & 22368$\pm$224         \\   
             &             &              &       &         \\
             &             &              &       &         \\
APMCC 815    &             &              &       &         \\
             &             &              &       &         \\
8447.0322    & 22 46 08.4 & -45 58 32.0 & 15.19 & 26954$\pm$245         \\   
             & 22 47 07.9 & -45 54 58.0 & 15.26 & 15578$\pm$154         \\   
             & 22 46 59.4 & -45 57 03.9 & 16.56 & 15235$\pm$154         \\   
             & 22 47 10.1 & -45 58 17.8 & 16.96 & 24534$\pm$246         \\   
             & 22 45 57.9 & -45 46 00.1 & 16.96 & 36441$\pm$191         \\   
             & 22 47 01.8 & -45 59 51.9 & 17.36 & 35620$\pm$341         \\   
             & 22 47 12.5 & -45 52 27.6 & 17.46 & 36578$\pm$177         \\   
             & 22 45 33.1 & -45 58 29.8 & 17.46 & 26620$\pm$176         \\   
             &             &              &       &         \\
APMCC 824    &             &              &       &         \\
             &             &              &       &         \\
8453.0439    & 22 50 05.9 & -51 44 08.0 & 14.21 & 28884$\pm$223         \\   
             & 22 50 49.3 & -51 44 21.5 & 17.13 & 12740$\pm$265         \\    
             & 22 49 41.7 & -51 44 43.4 & 17.43 & 29634$\pm$226         \\   
             & 22 50 05.3 & -51 34 25.9 & 17.43 & 28876$\pm$223         \\   
             & 22 49 30.8 & -51 46 57.2 & 17.53 & 28644$\pm$201         \\   
             & 22 49 58.5 & -51 47 32.5 & 17.73 & 30226$\pm$432         \\   
             & 22 49 32.3 & -51 44 36.8 & 17.73 & 29086$\pm$353         \\   
             & 22 49 25.5$^*$ & -51 44 36.0 &  --   & 17871$\pm$480         \\  
             & 22 49 57.0$^*$ & -51 47 48.0 &  --   & 30886$\pm$152         \\ 
             & 22 49 59.0$^*$ & -51 48 54.0 &  --   & 30550$\pm$169         \\ 
             & 22 49 57.0$^*$ & -51 49 27.0 &  --   & 29840$\pm$249         \\ 
             & 22 49 55.0$^*$ & -51 49 18.0 &  --   & 16630$\pm$382         \\ 
             &             &              &       &         \\
APMCC 864    &             &              &       &         \\
             &             &              &       &         \\
             & 23 11 36.2 & -28 59 53.4 & 17.22 & 31076$\pm$173         \\
             & 23 12 17.6 & -29 09 36.7 & 17.52 & 30902$\pm$241         \\
             &             &              &       &         \\
APMCC 898    &             &              &       &         \\
             &             &              &       &         \\
             & 23 26 38.0 & -23 46 04.0 & 17.33 & 37686$\pm$276         \\   
             & 23 27 12.5 & -23 44 49.0 & 17.83 & 34467$\pm$228         \\   
             & 23 26 45.0 & -23 56 56.8 & 18.13 & 32545$\pm$177         \\   
\end{tabular}
\end{minipage}
\end{table*}

\begin{table*}
\begin{minipage}{80mm}
\contcaption{}
\tabskip =1em plus2em minus.5em
\begin{tabular}{llllr}
Name  & $\alpha$(2000) & $\delta$(2000) & $b_j$ & V$_r$ \\
\noalign {\vskip 10pt}
\noalign {\hrule}
\noalign {\vskip 10pt}
             &             &              &       &         \\
APMCC 915    &             &              &       &         \\
             &             &              &       &         \\
8456.0477    & 23 39 28.8 & -45 57 59.0 & 12.70 & 19186$\pm$375         \\   
8456.0637    & 23 40 00.8 & -45 52 33.0 & 15.20 & 17477$\pm$184         \\   
             & 23 39 51.5 & -45 59 39.0 & 17.20 & 15942$\pm$206         \\   
             & 23 39 24.6 & -46 03 17.8 & 17.20 & 20648$\pm$293         \\   
             & 23 40 15.9 & -45 52 03.5 & 17.30 & 17293$\pm$169         \\   
             & 23 40 10.9 & -45 59 38.0 & 17.30 & 20939$\pm$198         \\   
             & 23 39 30.8 & -45 53 43.8 & 17.40 & 20654$\pm$196         \\   
             & 23 39 37.2 & -46 02 25.4 & 17.70 & 19155$\pm$293         \\   
             & 23 39 21.5 & -46 02 35.9 & 17.80 & 21298$\pm$261         \\   
             & 23 39 30.2 & -46 00 56.4 & 17.80 & 19629$\pm$226         \\   
             & 23 39 54.5 & -46 00 48.6 & 17.80 & 21222$\pm$138         \\   
             & 23 39 53.3 & -46 00 11.9 & 17.80 & 33921$\pm$380         \\   
\end{tabular}
\end{minipage}
\end{table*}

\begin{table*}
\begin{minipage}{80mm}
\caption{New velocity dispersions in APM clusters. ($\dag$) Velocity data 
from Las Campanas Redshift Survey}
\tabskip =1em plus2em minus.5em
\label {table4}
\begin{tabular}{llrr}
Cl. Id. & $<$z$_{new}>$ & $\sigma$ & N \\
 \noalign {\vskip 10pt}
 \noalign {\hrule}
 \noalign {\vskip 10pt}

APMCC 050 & 0.064 & 693$\pm$251 &  7 \\
APMCC 073 & 0.112 & 867$\pm$260 &  7 \\
APMCC 132 & 0.067 & 1123$\pm$233 &  12 \\
APMCC 160 & 0.076 & 573$\pm$285 &  7 \\
APMCC 352($\dag$) & 0.083 & 795$\pm$120 & 18 \\  
APMCC 359 & 0.067 & 399$\pm$180 &  9 \\
APM 031451.8-510556 & 0.076 & 748$\pm$144 & 10 \\
APMCC 369 & 0.075 & 700$\pm$ 98 & 29 \\
APMCC 400 & 0.071 & 528$\pm$232 &  8 \\
APM 221539.0-390817($\dag$) & 0.142 & 729$\pm$142 & 22 \\  
APMCC 815 & 0.090 & 425$\pm$158 &  8 \\
APMCC 824 & 0.098 & 780$\pm$127 & 11 \\
APMCC 915 & 0.068 & 751$\pm$179 & 10 \\
\end{tabular}
\end{minipage}
\end{table*}

\begin{table*}
\begin{minipage}{80mm}
\caption{Comparison of Velocity dispersions with Fadda et al. (1996). ($\dag$) Velocity data
from Las Campanas Redshift Survey}
\tabskip =1em plus2em minus.5em
\label {table5}
\begin{tabular}{llrrrr}
Cl. Id. & $<$z$_{new}> $ & $\sigma$ & N & $\sigma$(lit) & N \\
 \noalign {\vskip 10pt}
 \noalign {\hrule}
 \noalign {\vskip 10pt}
APMCC 015 & 0.061 & 652$\pm$248 &  9 & 628$\pm$61 & 80 \\
APMCC 042($\dag$) & 0.098 & 790$\pm$167 & 17 & 768$\pm$139 & 20 \\
APMCC 173($\dag$) & 0.080 & 416$\pm$ 87 & 19 & 547$\pm$159 & 30 \\  
APMCC 746($\dag$) & 0.072 & 509$\pm$127 & 12 & 677$\pm$141 & 29 \\  
\end{tabular}
\end{minipage}
\end{table*}


\begin{thebibliography}{}

 
\bibitem [Abell 1958] {abell1}  Abell, G. O. 1958, ApJS, 3, 211
\bibitem [Abell et al. 1989] {abell2} Abell, G. O., Corwin, H. G., Jr. \& Olowin,
R. P. 1989, ApJS, 70, 1 
\bibitem [A&M]{A&M} Arp, H.C. and Madore, B.F. 1987. A Cataloge of Southern Peculiar 
Galaxies and Associations. Cambridge University Press.
\bibitem [Bahcall and Cen 1993] {B&C} Bahcall, N. A.; Cen, R. 1993. ApJ Lett. 407, 49.
\bibitem [1990] {bee90} Beers, T. C., Flynn, K. \& Gebhardt, K. 1990, AJ, 100, 32
\bibitem [Collins et al. 1995] {collins} Collins, C. A., Guzzo, L., Nichol, R. C. 
\& Lumsden, S. L. 1995, MNRAS, 274, 1071
\bibitem [1992] {daltonth} Dalton G. 1992. D.Phil thesis. Oxford University.
\bibitem [1994] {dalton} Dalton, G. B., Efstathiou, G., Maddox, S. J. \&
Sutherland, W. J. 1994, MNRAS, 269, 151
\bibitem [Di Nella et al. 1996] {dinella} Di Nella, H., Couch, W. J., Paturel,
 G. \& Parker, Q. A. 1996, MNRAS, 283, 367
\bibitem [Ebeling \& Maddox 1995] {ebeling}Ebeling, H. \& Maddox, S. J. 1995, MNRAS, 275, 1155
\bibitem [Eke et al. 1996] {eke96} Eke V.R, Cole S., Frenk C. 1996. MNRAS, 282, 263.
\bibitem [1996] {fadda} Fadda, D., Girardi, M., Giuricin, G., Mardirossian,
F. \& Mezzetti, M. 1996, ApJ, 473, 670
\bibitem [Girardi et al.] {gir93} Girardi, M.; Biviano, A.; Giuricin, G.;
                  Mardirossian, F.; Mezzetti, M. 1993. ApJ 404, 38
\bibitem [1989] {cosmos} Heydon-Dumbleton, N. H., Collins, C. A. \& MacGillivray, H. T. 1989, 
MNRAS, 238, 379
\bibitem [Katgert et al. 1998] {kat98} Katgert P., Mazure A., den Hartog R., 
Adami C., Biviano A., Perea J. 1998. Astronomy \& Astrophysics Supp. 129, 399.
\bibitem [Lasket et al. 1990] {lasker} Lasker, B. M., Sturch, C. R., McLean, B. M., Russel, J. L.,
 Jenker, H. \& Shara, M. 1990, AJ, 99, 2019
\bibitem [Loveday 1996] {loveday} Loveday, J. 1996, MNRAS, 278, 1025
\bibitem [Loveday et al. 1996] {lovedayetal} Loveday, J., Peterson, 
B. A., Maddox, S. J. \& Efstathiou, G. 1996, ApJS, 107, 201
\bibitem [1992] {lumsden} Lumsden, S. L., Nichol, R. C., Collins, C. A. \& Guzzo, L. 1992, MNRAS 258, 1
\bibitem [Maddox et al. 1990a] {maddoxa} Maddox, S. J., Sutherland, W. J., Efstathiou, G. \&
Loveday, J. 1990a, MNRAS, 243, 692
\bibitem [Maddox et al. 1990b] {maddoxb} Maddox, S. J., Efstathiou, G. \& Sutherland, 
W. J. 1990b, MNRAS, 246, 433
\bibitem [Mazure et al. 1996]{maz} Mazure A., Katgert P., Den Hartog R., Biviano A.,
Dubath P., Escalera E., Focardi P., Gerbal D.,
Giuricin G., Jones B., Le Fevre O., Moles M., Perea J.,
Rhee G. 1996. A\& A, 310, 31.
\bibitem [Muriel et al. 1999] {mur99} Muriel H. et al. 1999. In preparation.
\bibitem [Paturel 1989] {paturel} Paturel, G. 1989. Catalog of Principal Galaxies (PGC). Lyon:
Base de Donnees Extragalactiques, Observatoire de Lyon, lc1980.
\bibitem [Paturel et al. 1995] {leda} Paturel, G., Vauglin, I., Andernach, H., Garnier, R.,
Marthinet, M., Petit, C., Di Nella, H., Bottinelli, L., Gouguenheim, L. \& Durand, N. 1995.
Information \& On-line Data in Astronomy, eds. D. Egret \& M. A.
Albrecht, Kluwer Acad. Publ., p. 115-126, Astrophysics and Space Science Library 203. 
LEDA.
\bibitem [Schectman et al. 1996] {schectman} Shectman, S. A., Landy, S. 
D., Oemler, A., Tucker, D. L., Lin, H., Kirshner, R. P. \& Schechter, P. L. 1996, ApJ, 470, 172 
\bibitem [Sutherland 1988] {suther} Sutherlanid W. 1988. MNRAS. 234, 159.
\bibitem [1979] {tonry} Tonry, J. L. \& Davis, M. 1979, AJ, 84, 1511
\bibitem [van Haarlem et al. 1997] {vanhaarlem} van Haarlem, M. P., Frenk, C. S. 
\& White, D. M. 1997, MNRAS, 287, 817
\bibitem [ZabLudoff et al. 1993] {zab} Zabludoff A. I., Geller M. J.,
                  Huchra J. P., Ramella M. 1993. AJ 106, 1301.

\end{thebibliography}
\end{document}